\documentclass[aps,prd,twocolumn,nofootinbib]{revtex4-1}

\usepackage{graphicx}

\newcommand{\bc}{\begin{center}}
\newcommand{\ec}{\end{center}}

\begin{document}

\title{Revisiting the Twist-3 Distribution Amplitudes of $K$ Meson within the QCD Background Field Approach}

\author{Tao Zhong$^{1}$}
\author{Xing-Gang Wu$^{1,2}$}
\email[email: ]{wuxg@cqu.edu.cn}
\author{Hua-Yong Han}
\author{Qi-Li Liao}
\author{Hai-Bing Fu}
\author{Zhen-Yun Fang}

\address{$^{1}$ Department of Physics, Chongqing University, Chongqing 401331, P.R. China\\
$^{2}$ SLAC National Accelerator Laboratory, 2575 Sand Hill Road, Menlo Park, CA 94025, USA}

\date{\today}

\begin{abstract}

In the present paper, we investigate the kaon twist-3 distribution amplitudes (DAs) $\phi_{p,\sigma}^K$ within the QCD background field approach. The $SU_f(3)$-breaking effects are studied in detail under a systematical way, especially the sum rules for the moments of $\phi_{p,\sigma}^K$ are obtained by keeping all the mass terms in the $s$-quark propagator consistently. After adding all the uncertainties in quadrature, the first two Gegenbauler moments of $\phi_{p,\sigma}^K$ are $a^1_{K,p}(1 {\rm GeV}) = -0.376^{+0.103}_{-0.148}$, $a^2_{K,p}(1 {\rm GeV}) = 0.701^{+0.481}_{-0.491}$, $a^1_{K,\sigma}(1 {\rm GeV}) = -0.160^{+0.051}_{-0.074}$ and $a^2_{K,\sigma}(1 {\rm GeV}) = 0.369^{+0.163}_{-0.149}$, respectively. Their normalization parameters $\mu_K^p |_{1\rm GeV} = 1.188^{+0.039}_{-0.043}$ GeV and $\mu_K^\sigma |_{1\rm GeV} = 1.021^{+0.036}_{-0.055}$ GeV. A detailed discussion on the properties of $\phi^K_{p,\sigma}$ moments  shows that the higher-order $s$-quark mass terms can indeed provide sizable contributions. Furthermore, based on the newly obtained moments, a model for the kaon twist-3 wavefunction $\Psi_{p,\sigma}^K(x,\mathbf{k}_\perp)$ with a better end-point behavior is constructed, which shall be useful for perturbative QCD calculations. As a byproduct, we make a discussion on the properties of the pion twist-3 DAs. \\

\noindent {\bf PACS numbers:} 12.38.Aw, 14.40.Df, 11.55.Hx

\noindent {\bf Key words:} QCD sum rules, Twist-3 wavefunction, Kaon

\end{abstract}

\maketitle

\section{introduction}

Meson distribution amplitude (DA), which describes the momentum fraction distribution of the parton in meson, is an important component for the QCD light-cone sum rule (LCSR) and the QCD factorization theory \cite{LCSR1,LCSR2,excu1,excu2,excu3}. In dealing with the exclusive processes, it is convenient to arrange the meson's DA by its different twist structures. The leading-twist DA shows the momentum distribution of the valence quarks in the meson, which usually provides major contribution to the QCD exclusive processes. The higher-twist DAs describe either the contributions from the higher Fock states with additional gluons and / or quark-antiquark pairs or the contributions from the transverse motion of quarks (antiquarks) in the leading-twist components. Usually, the contributions from the higher-twist DAs are power suppressed to that of the leading-twist in the large $Q^2$-region. However, the twist-3 DAs may provide sizable contributions for certain cases, so it arouses people's more and more interests, c.f. Refs.\cite{T3_85,T3_90,T3_98_PBall,T3_99_PBall,T3_04_HT,T3_05_HT,T3_11_ZHONG,T3_K,twist31,PI_T3_MODEL}.

Kaon twist-3 DAs are important input parameters for the kaon electromagnetic form factor, the $B \to K$ transition form factor and etc., whose properties have been investigated within the QCD sum rules and the $k_T$ factorization approach accordingly \cite{scale,BK_ff,T3_05_HT,T3_99_PBall,T3_K,twist31,KE}. More precise data are coming at LHC, it would be useful to study the higher-order and higher-power suppressed contributions so as to provide a deeper understanding of standard model parameters. For example, it has been pointed out that the $SU_f(3)$-breaking effect is about $10\%$ for $B\to K$ transition form factors, so a careful study on $K$ meson distributions shall lead to a better estimation of these form factors.

We have studied the QCD sum rules for the pionic twist-3 DAs in Ref.\cite{T3_11_ZHONG}, which are based on the framework of the QCD background field theory \cite{BG1,BG2,BG3,BG4,BG_HT,pro_func}. In the present paper, we shall improve our technology adopted there and then investigate the twist-3 DAs of $K$ meson by carefully dealing with its $SU_f(3)$-breaking effect. Such an effect is responsible for the different behaviors between kaon and pion DAs.

Basic assumption of the QCD sum rules is the introducing of the nonvanishing vacuum condensates such as the quark condensate $\left<\bar{q}q\right>$ and the gluon condensate $\left<G^2\right>$. Different to the conventional SVZ sum rules \cite{svz}, the background field approach provides a systematic description for these vacuum condensates from the viewpoint of field theory. And, it is convenient to derive useful relations among different non-perturbative matrix elements. Under the background field approach, it assumes that the quark and gluon fields are composed of the background fields and the quantum fluctuations around them. Nonperturbative effects can be described by the vacuum expectation values of these background fields, while the calculable perturbative effects are expressed by quantum fluctuations. Then to take the background field theory as the theoretical foundation for the QCD sum rules, it not only has distinct physical picture, but also can greatly simplify the calculation due to its capability of adopting different gauge conditions for quantum fluctuations and background fields respectively.

Because of the influence from background fields, the quark and gluon propagators shall include nonperturbative component inevitably. For the SVZ sum rules, one usually takes the following quark propagator formula \cite{pro1}
\begin{eqnarray}
S(x,0) &=& \frac{i \not\! x}{2\pi^2 x^4} - \frac{1}{12} \left<\bar{q}q\right> - \frac{x^2}{192} \left<g_s\bar{q} \sigma T G q\right> - \frac{m}{4\pi^2 x^2} \nonumber\\
&& + \frac{i m}{48} \left<\bar{q}q\right> \not\! x  + \frac{i m x^2}{1152} \left<g_s\bar{q}\sigma T G q\right> \not\! x +\cdots ,
\label{slqp}
\end{eqnarray}
where $\cdots$ stands for even higher-dimensional terms and higher-order mass terms. Note that the above quark propagators in configuration space are given as an expansion in quark mass, and the mass terms are kept only up to first order. For the light-quark propagators, it is enough. However, the omitted higher-order mass terms may lead to sizable contributions to the meson or baryon with heavy quark(s). Even for the case of $K$ meson, the contributions from higher-order $s$-quark mass terms, either positive or negative, are sizable. Hence to obtain a better understanding of $SU_f(3)$-breaking effect for $K$ meson, one needs to take the sizable mass terms into consideration in a more proper way.

The remaining parts of the paper are organized as follows. In Sec.II, we present the calculation technology for deriving the sum rules for the moments of the kaon twist-3 DAs. And a model for the kaon twist-3 wave functions is also presented. Numerical results are given in Sec.III, where the properties of kaon twist-3 DAs are discussed. Sec.IV is reserved for a summary. In the Appendix, we give useful formulas for simplifying the matrix elements $\left<0\left| \bar{\psi}^a_\alpha (x) \psi^b_\beta (y) \right|0\right>$ and $\left<0\left| \bar{\psi}^a_\alpha (x) \psi^b_\beta (y) G^A_{\mu\nu} \right|0\right>$.

\section{Calculation technology}

\subsection{Sum rules for the pseudoscalar twist-3 DAs}

Under the background field theory, the quark and gluon propagators satisfy the following equations \cite{pro_func}:
\begin{equation}
(i \not\! D - m)S_F(x,0) = \delta^4(x)
\end{equation}
and
\begin{equation}
\left( g^{\mu\nu}(\widetilde{D}^2)^{ab} + 2 f^{abc}G^{c\mu\nu} \right) S^{bd}_{\nu\rho}(x,0) = \delta^{ad}g^\mu_{\ \rho}\delta^4(x) ,
\end{equation}
where $D_\mu = \partial_\mu - igT^a A^a_\mu$ and $(\widetilde{D}_\mu)^{ab} = \delta^{ab}\partial_\mu - g f^{abc} A^{c}_\mu$ are gauge covariant derivatives in the fundamental and adjoint representations respectively. $A^a_\mu$ is the background gluon field, $G^a_{\mu\nu} = \partial_\mu A^a_\nu - \partial_\nu A^{a}_\mu + g f^{abc} A^b_\mu A^c_\nu$ is the gluon field strength tensor and $f^{abc} (a,b,c = 1,2,\cdots ,8)$ is the structure constant of $SU(3)$ group. Fixing the gauge freedom of the background field by the fixed-pointed gauge \cite{fixed1,fixed2}, $x^\mu A^a_\mu = 0$, we obtain
\begin{widetext}
\begin{eqnarray}
S_{F}(x,0) = i \int \frac{d^4 q}{(2\pi)^4} e^{-iq \cdot x} \left\{ -\frac{m + \not\! q}{m^2 - q^2} + \frac{\gamma^\nu ( \not\! q - m ) \gamma^\mu}{(m^2 - q^2)^2} b_{0\nu\mu} - i \left[ 2 \frac{\gamma^\nu (\not\! q - m) q^\rho}{(m^2 - q^2)^3} +  \frac{g^{\nu\rho}}{(m^2 - q^2)^2} \right] \gamma^\mu b_{1\nu\mu|\rho} +\cdots \right\} \label{prop of s3}
\end{eqnarray}
\end{widetext}
for the quark propagator and
\begin{equation}
S^{ab}_{\mu\nu}(x,0)=i \int \frac{d^4 q}{(2\pi)^4} e^{-iq \cdot x} \left\{-\frac{g_{\mu\nu}}{q^2} \delta^{ab} +\cdots \right\}
\end{equation}
for the gluon propagator, where $b_{0\nu\mu}=\frac{i}{2}G_{\nu\mu}(0)$ and $b_{1\nu\mu|\rho} = \frac{i}{3} \left[ G_{\nu\mu;\rho}(0)+ G_{\rho\mu;\nu}(0)\right]$. Here $G_{\nu\mu}(x) = g_s T^a G^a_{\nu\mu}(x)$, the gauge invariant function $G_{\nu\mu;\rho}(0)=g_s T^a \widetilde{D}^{ab}_{\rho} G^{b}_{\nu\mu}(x)|_{x=0}$, and the symbol $\cdots$ stands for the irrelevant terms for our present analysis that will lead to higher-order operators over dimension-six. To use a propagator in momentum-space form as Eq.(\ref{prop of s3}) has been suggested in the literature already, e.g. it has been suggested to deal with the $D^*D\pi$ and $B^*B\pi$ couplings in Ref.\cite{ddpi}. However in these discussions, usually the first two terms in the quark-propagator are kept only. For the present case, one may find that the third term should be kept to provide a more accurate sum rules up to dimension-six operators. As a special case, by taking only the first-order mass term, we can obtain the quark propagator in the coordinate space,
\begin{eqnarray}
S(x,0) &=& \frac{i \not\! x}{2\pi^2 x^4}  - \frac{\gamma^\alpha \not\! x \gamma^\beta}{16\pi^2 x^2} G_{\alpha\beta}(0) + \frac{\ln (-x^2)}{48\pi^2} \gamma^\mu G_{\alpha\mu;}^{\ \ \ \alpha}(0) \nonumber\\
&&\!\!\!\!\!\! - \frac{x^\alpha \gamma^\nu \not\! x \gamma^\mu}{48\pi^2 x^2} \left[ G_{\nu\mu;\alpha}(0) + G_{\alpha\mu;\nu}(0)\right] - \frac{m}{4\pi^2 x^2} + \cdots . \label{lqp}
\end{eqnarray}
Since the quark propagator in momentum space keeps the mass terms naturally, so we shall adopt (\ref{prop of s3}) other than (\ref{lqp}) to do the following calculation. In fact, as will be shown later, the high-order mass terms are indeed important for giving a more sound $SU_f(3)$-breaking effect in $\phi^K_{p,\sigma}$.

The pseudoscalar twist-3 DAs $\phi_p^P$ and $\phi_\sigma^P$ are defined as,
\begin{eqnarray}
\left<0\left| \bar{q}_1(z)i\gamma_5q_2(-z) \right|P(q)\right> = f_P \mu_P^p \int^1_0 du \phi^P_p(u) e^{i\xi(z\cdot q)} , \nonumber
\end{eqnarray}
\begin{eqnarray}
&&\left<0\left| \bar{q}_1(z)\sigma_{\mu\nu}\gamma_5q_2(-z) \right|P(q)\right> = \nonumber\\
&& \quad\quad -i(q_\mu z_\nu - q_\nu z_\mu) \frac{1}{3} f_P \mu_P^\sigma \int^1_0 du \phi^P_\sigma(u) e^{i\xi(z\cdot q)} , \nonumber
\end{eqnarray}
where $\xi = 2x-1$, $q_1 = d$ and $q_2 = u$ for pion, $q_1=s$ and $q_2=u$ for kaon, respectively; the parameters $f_P$ and $\mu_P^{p,\sigma}$ stand for the decay constant and the normalization parameter of the pseudoscalar, respectively. The DA moments are defined as
\begin{eqnarray}
\left<\xi^n_p\right>_P &=& \int^1_0 dx (2x-1)^n \phi_p^P(x) , \label{moment_definition0} \\
\left<\xi^n_\sigma\right>_P &=& \int^1_0 dx (2x-1)^n \phi_p^\sigma(x), \label{moment_definition}
\end{eqnarray}
which satisfy
\begin{eqnarray}
\left<0\left| \bar{q}_1(0)\gamma_5 (iz\cdot \tensor{D})^n q_2(0) \right|P(q)\right> =  -if_P \mu_P^p \left<\xi^n_p\right>_P (z\cdot q)^n   \nonumber
\end{eqnarray}
and
\begin{eqnarray}
&& \left<0\left| \bar{q}_1(0)\sigma_{\mu\nu}\gamma_5 (iz\cdot \tensor{D})^{n+1} q_2(0) \right|P(q)\right> =  \nonumber\\
&&\quad -\frac{n+1}{3}f_P \mu_P^\sigma \left<\xi^n_\sigma\right>_P (q_\mu z_\nu - q_\nu z_\mu)(z\cdot q)^n
\nonumber
\end{eqnarray}
respectively. In deriving the sum rules for the moments, we adopt the following correlation functions:
\begin{widetext}
\begin{eqnarray} &&
(z\cdot q)^n I^{(n,0)}_{P,p} (q^2) \equiv -i \int d^4 x e^{iq\cdot x} \left<0\left| T \left\{ \bar{q}_1(x) \gamma_5 (iz\cdot \tensor{D})^n q_2(x), \bar{q}_2(0) \gamma_5 q_1(0) \right\} \right|0\right>
\label{cor-p}
\end{eqnarray}
\begin{eqnarray} &&
-i (q_\mu z_\nu - q_\nu z_\mu) (z\cdot q)^n I^{(n,0)}_{P,\sigma} (q^2) \equiv -i \int d^4 x e^{iq\cdot x} \left<0\left| T \left\{ \bar{q}_1(x) \sigma_{\mu\nu} \gamma_5 (iz\cdot \tensor{D})^{(n+1)} q_2(x), \bar{q}_2(0) \gamma_5 q_1(0) \right\} \right|0\right>.
\label{cor-si}
\end{eqnarray}
\end{widetext}

\begin{figure}
\begin{center}
\includegraphics[width=0.48\textwidth]{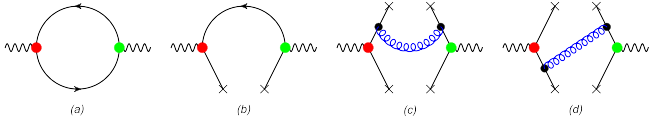}
\end{center}
\caption{Feynman diagrams for the pseudoscalar DA moments, where the background gluon fields are included in the Fermion propagators implicitly and the background quark fields are depicted as crosses. The left big dot stands for the vertex $\gamma_5 (i z\cdot \tensor{D}^n)$ and $\sigma_{\mu\nu} \gamma_5 (i z\cdot \tensor{D}^{(n+1)})$ for $\phi^P_p$ and $\phi^P_\sigma$ respectively.}
\label{tree}
\end{figure}

Fig.(\ref{tree}) shows the Feynman diagrams for the pseudoscalar DA moments, where the background gluon fields are included in the Fermion propagators implicitly and the background quark fields are depicted as crosses.

Following the standard QCD sum rule technology, the sum rules of the moments can be derived. And we obtain
\begin{widetext}
\begin{eqnarray}
&&\frac{1}{M^2} \left<\xi^n_p\right>_P f_P^2 (\mu^{p}_P)^2 e^{-\frac{m_P^2}{M^2}} \nonumber\\
&&\quad = \int^1_0 dx e^{-\frac{m_1^2}{M^2(1-x)}} \left\{ \frac{3}{4\pi^2} (2x-1)^n \left[ (n+3)M^2x(1-x) + m_1^2 x \right] + \left<\alpha_sG^2\right> \left( \frac{n+1}{M^2}+\frac{m_1^2}{M^4(1-x)}\right)\right. \nonumber\\ && \left. \times \left[ \frac{n(n-1)}{12\pi}(2x-1)^{n-2}x(1-x) + \frac{1}{8\pi}(2x-1)^n \right] - \left<g^3_s f G^3\right> \frac{n(n-1)}{96\pi^2}(2x-1)^{n-2}\frac{1}{M^4} \right\} \nonumber\\
&& - \frac{3}{4\pi^2} \int^1_0 dx (2x-1)^n \left[ (n+3)M^2x(1-x) \left( 1+\frac{s^p_P}{M^2} \right) + (n+2) m_1^2 x \right] e^{-\frac{s^p_P}{M^2}} \nonumber\\
&& + \left<\bar{q}_1q_1\right> \left[\frac{(n+1)m_1}{2M^2} + \frac{n(2n+1)}{6}\frac{m_1^3}{M^4} \right] + (-1)^n\left<\bar{q}_2q_2\right> \left[ -\frac{m_1}{M^2} + \frac{m_1^3}{M^4} \right] + \left<g_s\bar{q}_1\sigma TGq_1\right>\frac{n(5-8n)m_1}{36M^4}\nonumber\\
&&  + (-1)^n \left<g_s\bar{q}_2\sigma TGq_2\right> \left[ \frac{18(n-1)m_1}{36M^4} + \frac{(3-2n)m_1^3}{4M^6} \right]+ \left[ \left<g_s\bar{q}_1q_1\right>^2 + (-1)^n \left<g_s\bar{q}_2q_2\right>^2 \right] \frac{2n^2+7n-12}{81M^4}  \nonumber\\
&& - (-1)^n \left<g_s\bar{q}_2q_2\right>^2 \frac{2n^2+7n-15}{81} \frac{m^2_1}{M^6} + 4\pi\alpha_s \left<\bar{q}_1q_1\right>\left<\bar{q}_2q_2\right> \left\{ -\frac{2}{9}\frac{m_1^2}{M^6} + \frac{2}{9}\left[ 1+(-1)^n \right] \left(\frac{1}{M^4} - \frac{m_1^2}{2M^6}\right) \right\}  \label{srkp}
\end{eqnarray}
and
\begin{eqnarray}
&&\frac{1}{3M^2} \left<\xi^n_\sigma\right>_P f_P^2 (\mu_P^{\sigma})^2 e^{-\frac{m_P^2}{M^2}}
\nonumber\\
&&\quad = \int^1_0 dx e^{-\frac{m^{2}_1}{M^2(1-x)}} \left\{ \frac{3}{4\pi^2}(2x-1)^n M^2 x(1-x) + \left<\alpha_s G^2\right> \left[\frac{n(n-1)}{12\pi}\frac{(2x-1)^{n-2}x(1-x)}{M^2}+ \frac{(2x-1)^n}{24\pi M^2} \right] \right\}  \nonumber\\
&&  - \frac{3}{4\pi^2} \int^1_0 dx (2x-1)^n \left[ M^2 x(1-x) \left( 1+\frac{s^\sigma_P}{M^2} \right) + m_1^2 x \right] e^{-\frac{s^\sigma_P}{M^2}} + \left<\bar{q}_1q_1\right> \left[ \frac{m_1}{2M^2} + \frac{(2n+1)m_1^3}{6M^4} \right] \nonumber\\
&& - \left<g_s \bar{q}_1\sigma T G q_1\right> \frac{(8n+1)m_1}{36M^4} - (-1)^n\left<g_s\bar{q}_2\sigma T G q_2\right> \left[ \frac{m_1}{6M^4} - \frac{m_1^3}{6M^6} \right] + \left[ \left<g_s\bar{q}_1q_1\right>^2 + (-1)^n \left<g_s\bar{q}_2q_2\right>^2 \right] \frac{2n-5}{81M^4} \nonumber\\ &&
- (-1)^n \left<g_s\bar{q}_2q_2\right>^2 \frac{(2n-3)m_1^2}{81M^6} .
\label{srksi}
\end{eqnarray}
\end{widetext}
Because the current quark mass of $u(d)$-quark is quite small, we have set $m_2=m_u \simeq 0$, $m_1 =m_d \simeq 0$ for pion and $m_1 =m_s$ for kaon accordingly. $M$ stands for the Borel parameter, $m_P$ is the pseudoscalar mass, and $s^{p,\sigma}_P$ are continuum threshold. The non-perturbative matrix elements: $\left<\alpha_s G^2\right> = \left<\alpha_s G^A_{\mu\nu} G^{A\mu\nu}\right>$, $\left<g_s\bar{q}\sigma T G q\right> = \left<g_s\bar{q} \sigma_{\mu\nu} T^A G^{A\mu\nu}q\right>$, $\left<g^3_s f G^3\right> = \left<g_s^3 f^{ABC} G^A_{\mu\nu}G^{B\nu\rho}G^{C\mu}_\rho\right>$ \footnote{Note in the condensate $\left<g_s \bar{q}q\right>^2$, whose the coupling constant $g_s$ comes from the gluonic background field, so we should treat the condensate as a whole.}. If setting $n=0$ in Eqs.(\ref{srkp},\ref{srksi}), one can obtain the sum rules of the normalization parameters. In deriving the sum rules (\ref{srkp},\ref{srksi}), we have implicitly adopted the following Borel transformation formulas
\begin{eqnarray}
&&\hat{L}_M (AQ^2+B)^k \ln (AQ^2+B) \nonumber\\
&&=(-1)^{k+1} k! (AM^2)^k \exp\left(-\frac{B}{AM^2}\right) \quad (k\geq 0) , \nonumber
\end{eqnarray}
\begin{eqnarray}
\hat{L}_M \frac{1}{(m^2+Q^2)^k} = \frac{1}{(k-1)!} \frac{1}{M^{2k}} \exp \left(-\frac{m}{M^2}\right) \quad (k \geq 1) , \nonumber
\end{eqnarray}
and we have used the simplified matrix elements $\left<0\left| \bar{\psi}^a_\alpha (x) \psi^b_\beta (y) \right|0\right>$ and $\left<0\left| \bar{\psi}^a_\alpha (x) \psi^b_\beta (y) G^A_{\mu\nu} \right|0\right>$, whose detailed derivations are presented in the Appendix.

From the sum rules (\ref{srkp},\ref{srksi}), it is found that their perturbative parts and the dimension-four gluon condensate part come from Fig.(\ref{tree}a); the dimension-three quark-antiquark condensate part and the dimension-five quark-gluon condensate part come from Fig.(\ref{tree}b); and the dimension-six four-quark condensate part comes from Figs.(\ref{tree}b,\ref{tree}c,\ref{tree}d). Numerically, it can be found that the contribution for $\phi^P_p$ from Fig.(\ref{tree}c) is very small ($\sim m_1^2/M^6$). But if one adopts the quark propagator (\ref{lqp}), the contribution of $\phi^{P}_p$ from Fig.(\ref{tree}c) should be twice than that of Fig.(\ref{tree}d) ($\sim 1/M^4$) and is sizable. This shows that by keeping the mass-terms properly, one can obtain a correct estimation of the relative importance among different Feynman diagrams. To show the $s$-quark mass effects more clearly, we shall discuss the different consequences caused by the using of the propagators (\ref{prop of s3}) and (\ref{lqp}) in the next section. And we shall find the importance of using the propagator (\ref{prop of s3}), which keeps the higher order mass-terms in a more consistent way.

It is well-known that the kaon twist-3 DAs can be expanded in Gegenbauler polynomials as
\begin{eqnarray}
&& \phi^K_p(\mu_f,x) = 1 + \sum_{n=1}^\infty a_{K,p}^n(\mu_f) C^{1/2}_{n}(2x-1) ,
\label{phi_g_m0}\\
&& \phi^K_\sigma(\mu_f,x)=6x(1-x)\left[1 + \sum_{n=0}^\infty a_{K,\sigma}^n(\mu_f) C^{3/2}_{n}(2x-1)\right] , \label{phi_g_m1}
\end{eqnarray}
where $C^{1/2,3/2}_{n}(2x-1)$ are Gegenbauler polynomials, $a_{K,p}^n(\mu_f)$ and $a_{K,\sigma}^n(\mu_f)$ are Gegenbauler moments at the factorization scale $\mu_f$.

With the help of the above sum rules (\ref{moment_definition0}, \ref{moment_definition}) for the DA moments, we can obtain the kaon twist-3 DAs, e.g.
\begin{eqnarray}
&& a_{K,p}^1 = 3 \left<\xi_p^1\right>_K, \quad a_{K,p}^2 = \frac{15}{2} \left(\left<\xi_p^2\right>_K - \frac{1}{3}\right), \nonumber\\
&& a_{K,\sigma}^1 = \frac{5}{3} \left<\xi_\sigma^1\right>_K, \quad a_{K,\sigma}^2 = \frac{35}{12} \left(\left<\xi_\sigma^2\right>_K - \frac{1}{5}\right). \label{gegen_moment}
\end{eqnarray}
And the Gegenbauler moments at any scale $\mu$ can be obtained from the renormalization group equations from an initial factorization scale \cite{scale}:
\begin{eqnarray}
a_{K,p(\sigma)}^n(\mu) = a_{K,p(\sigma)}^n(\mu_f) \left(\frac{\alpha_s (\mu)} {\alpha_s (\mu_f)} \right)^{\gamma_n / \beta_0}, \label{rge}
\end{eqnarray}
where $\gamma_n = C_F \left( 1 - \frac{2}{(n+1)(n+2)} + 4 \sum^{n+1}_{m=2} \frac{1}{m} \right)$ and $\beta_0 = (11 N_c - 2 N_f)/3$ with $C_F=4/3$.

\subsection{kaon twist-3 wavefunctions}

The kaon wave function and its DA can be related with the following equation,
\begin{equation} \label{phips}
\phi^{K}_{p,\sigma} (x,\mu_f) = \int_{|\textbf{k}_\perp|<\mu_f} \frac{d^2 \textbf{k}_\perp}{16\pi^3} \psi^{K}_{p,\sigma} (x,\textbf{k}_{\perp}) ,
\end{equation}
where $\mu_f \sim {\cal O}(1GeV)$ is the factorization scale. Due to the renormalization group equation (\ref{rge}), the distribution amplitudes under different choice of $\mu_f \sim {\cal O}(1GeV)$ can be related with each other through evolution, which shall result in the same behaviors at the present considered accuracy \cite{bhld}. Hereafter for definiteness, we set $\mu_f =1 GeV$.

Following the same idea of Refs.\cite{BK_ff,KE,PI_T3_MODEL,bhla,bhlb,bhlc,bhld} where its transverse momentum dependence is constructed on the BHL-prescription \cite{BHL}, the kaon twist-3 wave functions can be constructed as
\begin{eqnarray}
&&\psi^K_p(x,\textbf{k}_\perp) = \left[ 1 + B^K_p C^{1/2}_1 (2x-1) + C^K_p C^{1/2}_2 (2x-1) \right] \nonumber\\
&&\times \frac{A^K_p}{x(1-x)}  \exp \left[ -\frac{1}{8\beta^{K 2}_p} \left( \frac{\widetilde{m}^2_q + \textbf{k}^2_\perp}{x} + \frac{\widetilde{m}^2_s + \textbf{k}^2_\perp}{1-x} \right) \right]
\end{eqnarray}
and
\begin{eqnarray}
&&\psi^K_\sigma(x,\textbf{k}_\perp) = \left[ 1 + B^K_\sigma C^{3/2}_1 (2x-1) + C^K_\sigma C^{3/2}_2 (2x-1) \right] \nonumber\\
&&\times\frac{A^K_\sigma}{x(1-x)} \exp \left[ -\frac{1}{8\beta^{K 2}_\sigma} \left( \frac{\widetilde{m}^2_q + \textbf{k}^2_\perp}{x} + \frac{\widetilde{m}^2_s + \textbf{k}^2_\perp}{1-x} \right) \right] .
\end{eqnarray}
$\widetilde{m}_{q,s}$ indicate the constituent quark mass, and their standard values are $\widetilde{m}_q \simeq 0.30$ GeV and $\widetilde{m}_s \simeq 0.45$ GeV. The parameters $A^{K}_{p,\sigma}$, $B^{K}_{p,\sigma}$, $C^{K}_{p,\sigma}$ and $\beta^{K}_{p,\sigma}$ can be determined by the average value of the transverse momentum $\left<\textbf{k}^2_\perp\right>^{K}_{p,\sigma}$ ($\approx (0.350 \rm GeV)^2$ \cite{transverse_momentum}),
\begin{eqnarray}
\left<\textbf{k}^2_\perp\right>^{K}_{p,\sigma} = \frac{\int dx d^2\textbf{k}_\perp |\textbf{k}^2_\perp| |\psi^{K}_{p,\sigma}(x,\textbf{k}_\perp)|^2}{\int dx d^2 \textbf{k}_\perp |\psi^{K}_{p,\sigma}(x,\textbf{k}_\perp)|^2},
\end{eqnarray}
the wave function normalization
\begin{eqnarray}
\int^1_0 dx \int_{\textbf{k}_\perp<\mu_f} \frac{d^2 \textbf{k}_\perp}{16 \pi^3} \psi^{K}_{p,\sigma} (x,\textbf{k}_\perp) = 1,
\end{eqnarray}
and the first two DA moments
\begin{eqnarray}
a_{K,p}^n(\mu_f) = \frac{\int^1_0 dx \phi^K_p (x,\mu_f) C^{1/2}_{n} (2x-1)}{\int^1_0 dx \left[ C^{1/2}_{n} (2x-1) \right]^2},
\end{eqnarray}
\begin{eqnarray}
a_{K,\sigma}^n(\mu_f) = \frac{\int^1_0 dx \phi^K_\sigma (x,\mu_f) C^{3/2}_{n} (2x-1)}{\int^1_0 dx 6x(1-x) \left[ C^{3/2}_{n} (2x-1) \right]^2}.
\end{eqnarray}

\section{numerical analysis}

\subsection{input parameters}

From the Particle Data Group \cite{PDG}, we take the current $s$-quark mass as $m_s(2\textrm{GeV})=100^{+30}_{-20}$ MeV; $\pi$ and $K$ meson masses $m_\pi=139.57018 \pm 0.00035$ MeV and $m_K=493.677 \pm 0.016$ MeV; the pion and kaon decay constants $f_\pi = 130.41 \pm 0.20$ MeV and $f_K = 156.1 \pm 0.8$ MeV. The vacuum condensates have been calculated and updated since 1979 \cite{svz}, c.f. Refs.\cite{pro1,VC80,VC90,VC20,REW,VC2010,VC2011}. We take the dimension-four and dimension-six condensates to be \cite{VC2011}: $\left<\alpha_sG^2\right>=(7.5 \pm 2.0) \times 10^{-2} \textrm{GeV}^4$, $\left<g^3_sfG^3\right> = (8.3 \pm 1.0) \textrm{GeV}^2 \times \left<\alpha_sG^2\right>$. And for the quark condensate and quark-gluon condensate we take \cite{VC2010}: $\left<\bar{u}u\right>(2\textrm{GeV})=-(0.254 \pm 0.015)^3 \textrm{GeV}^3$, $\left<\bar{d}d\right>=\left<\bar{u}u\right>$, $\left<\bar{s}s\right>/\left<\bar{u}u\right>=0.74 \pm 0.03$, $\left<g_s\bar{q}\sigma TG q\right>=m^2_0 \left<\bar{q}q\right>$ with $m^2_0 = 0.80 \pm 0.02 \textrm{GeV}^2$, and $\left<g_s\bar{q}q\right>^2 =(2.7^{+0.5}_{-0.4})\times 10^{-3} \textrm{GeV}^6$. The continuum threshold parameter is taken to be around the mass square of the first exciting state of the meson. Considering the first exciting states are $\pi (1300)$ and $K(1460)$ for pion and kaon respectively \cite{PDG}, we take $s^{p,\sigma}_\pi=1.69 \pm 0.10$ $\textrm{GeV}^2$ and $s^{p,\sigma}_K=2.13 \pm 0.10$ $\textrm{GeV}^2$. The leading order $\alpha_s$ is fixed by $\alpha_s(M_Z) = 0.1184 \pm 0.0007$ \cite{as} and the renormalization scale is taken as $M$.

\subsection{$\phi^K_{p,\sigma}(x,\mu_f)$ and $\psi^K_{p,\sigma}(x,\mathbf{k}_\perp)$}

To derive proper Borel windows for the sum rules of $\phi^K_{p,\sigma}(x,\mu_f)$, the criteria are to suppress the unwanted continuum contribution and the higher-dimensional contribution as much as possible so as to obtain more accurate results.

\begin{widetext}
\begin{center}
\begin{table}
\caption{The normalization parameters and the first two moments of $\phi^K_{p,\sigma}$. }
\begin{tabular}{|c|c|c|c||c|c|}
\hline \hline
$M^2(\rm GeV^2)$ & $[0.875,1.103]$ & $M^2(\rm GeV^2)$ & $[0.800,1.198]$ & $M^2(\rm GeV^2)$ & $[1.088,1.094]$\\
$(\mu_K^{p})^2(\rm GeV^2)$ & $1.287 - 1.517$ & $\left<\xi^1_p\right>_K$ & $-0.126 \pm 0.010$ & $\left<\xi^2_p\right>_K$ & $0.425 \pm 0.001$\\
\hline\hline
$M^2(\rm GeV^2)$ & $[0.968,1.034]$ & $M^2(\rm GeV^2)$ & $[1.103,1.267]$ & $M^2(\rm GeV^2)$ & $[0.825,1.117]$\\
$(\mu_K^p \mu_K^\sigma) (\rm GeV^2)$ & $1.178 - 1.247$ & $\left<\xi^1_\sigma\right>_K$ & $-0.094 \pm 0.001$ & $\left<\xi^2_\sigma\right>_K$ & $0.329 \pm 0.013$ \\
\hline \hline
\end{tabular}
\label{tsrK}
\end{table}

\begin{table}
\caption{Main uncertainties for the normalization parameters and the first two moments of $\phi^K_{p,\sigma}$, where the uncertainty of a particular parameter is obtained by fixing other parameters to be their center values.}
\begin{tabular}{|c| c c | c c | c c|}
\hline \hline
 & ~~$ (\mu_K^{p})^2 $~~ & ~~$ (\mu_K^p \mu_K^\sigma)$~~ & ~~$ \left<\xi^1_p\right>_K $~~ & ~~$ \left<\xi^2_p\right>_K $~~ & ~~$ \left<\xi^1_\sigma\right>_K $~~ & ~~$ \left<\xi^2_\sigma\right>_K $~~ \\
\hline \hline
$m_s$                                 & $^{-0.014}_{+0.003}$ & $^{-0.077}_{+0.044}$ & $^{-0.045}_{+0.029}$ & $^{-0.025}_{+0.014}$ & $^{-0.038}_{+0.024}$ & $^{-0.009}_{+0.003}$ \\ \hline
$s_K^{p,\sigma}$                      & $^{+0.049}_{-0.051}$ & $^{+0.050}_{-0.053}$ & $\pm 0.000$          & $\pm 0.014$          & $\pm 0.000$          & $^{+0.007}_{-0.008}$ \\ \hline
$\left<\alpha_sG^2\right>$            & $\pm 0.041$          & $\pm 0.039$          & $\mp 0.002$          & $\pm 0.044$          & $\mp 0.001$          & $\pm 0.033$ \\ \hline
$\left<\bar{s}s\right>$               & $^{-0.004}_{+0.003}$ & $^{-0.023}_{+0.019}$ & $^{-0.011}_{+0.009}$ & $^{-0.013}_{+0.011}$ & $^{-0.017}_{+0.015}$ & $^{-0.020}_{+0.017}$ \\  \hline
$\left<\bar{u}u\right>$               & $^{+0.018}_{-0.016}$ & $\mp 0.001$          & $^{-0.011}_{+0.010}$ & $^{+0.012}_{-0.011}$ & $\pm 0.000$          & $^{-0.001}_{+0.000}$ \\  \hline
$\left<g_s\bar{s}\sigma TGs\right>$   & $\pm 0.000$          & $^{+0.002}_{-0.001}$ & $\pm 0.001$          & $^{+0.006}_{-0.005}$ & $^{+0.008}_{-0.006}$ & $^{+0.022}_{-0.018}$ \\  \hline
$\mu_K^{p,\sigma}$                    & $ - $                & $ - $                & $^{+0.008}_{-0.010}$ & $^{-0.027}_{+0.033}$ & $^{+0.006}_{-0.009}$ & $^{-0.021}_{+0.032}$ \\  \hline
$\left<g_s^3fG^3\right>$              & $ - $                & $ - $                & $ - $                & $^{-0.017}_{+0.014}$ & $ - $                & $ - $ \\
\hline\hline
\end{tabular}
\label{uncertK}
\end{table}
\end{center}
\end{widetext}

First, we determine the normalization parameters $\mu_K^{p,\sigma}$. In Refs.\cite{T3_90,T3_99_PBall}, it is calculated by using the idea of the quark equation of motion (QEM). While it has been pointed out that the quarks inside the meson is not exactly on-shell \cite{T3_05_HT}, so the results in Refs.\cite{T3_90,T3_99_PBall} is only an approximation. As a notation, it is found that the three-particle twist-3 distributions $\phi_{3\pi,3K}$ can be related with $\phi^{P}_{p,\sigma}$ through the QEM \cite{T3_90,T3_99_PBall}. However due to the similar reason, we do not discuss the three-particle distributions with those relations in the present paper. A simple discussion on this point can be found in Ref.\cite{T3_05_HT}, where compatible results for $\ll\alpha_{3}\gg$ and $f_{3\pi}$ with those derived from QEM \cite{T3_85} have been obtained through proper consideration. At the present, by setting $n=0$ in the sum rules (\ref{srkp},\ref{srksi}), we can obtain the sum rules for $\mu_K^{p,\sigma}$. To set the Borel window for $\mu_K^{p,\sigma}$, we take the continuum contribution to be less than $40\%$, and the dimension-six condensate contribution to be less than $2\%$ for $\mu_K^{p}$ and $4\%$ for $\mu_K^{\sigma}$. The values of $(\mu_K^{p})^2$ and $(\mu_K^{p} \mu_K^{\sigma})$ and the their corresponding Borel windows are collected in Tab.\ref{tsrK}, which is obtained by setting all the input parameters to be their center values. Main uncertainties caused by the current quark mass $m_{s}$, the continuum threshold $s^{p,\sigma}_{K}$, the dimensional operators and etc. are collected in Tab.\ref{uncertK}. Other smaller uncertainties caused by the parameters as $\left<g_s\bar{u}\sigma TGu\right>$, $\left<g_s\bar{s}s\right>^2$, $\left<g_s\bar{u}u\right>^2$, $f_K$ and etc. are not presented. By taking all uncertainty sources into consideration, we obtain,
\begin{eqnarray}
\mu_K^p |_{1\rm GeV} &=& 1.188^{+0.039}_{-0.043} \;\;\textrm{GeV} ,\\
\mu_K^\sigma |_{1\rm GeV} &=& 1.021^{+0.036}_{-0.055} \;\;\textrm{GeV} ,
\end{eqnarray}
where the renormalization group equation of $\mu_K^p$ and $\mu_K^\sigma$ \cite{rge90,rge20} has been adopted to run its value from the scale $M$ to $1$ GeV. Note our value of $\mu_K^p |_{1\rm GeV}$ is different from the value obtained by the on-shell condition (i.e. $\mu_K^p |_{1\rm GeV}\simeq 1.424$ GeV \cite{T3_99_PBall}) by about $(17\pm 3)\%$.

\begin{widetext}
\begin{center}
\begin{figure}
\includegraphics[width=0.24\textwidth]{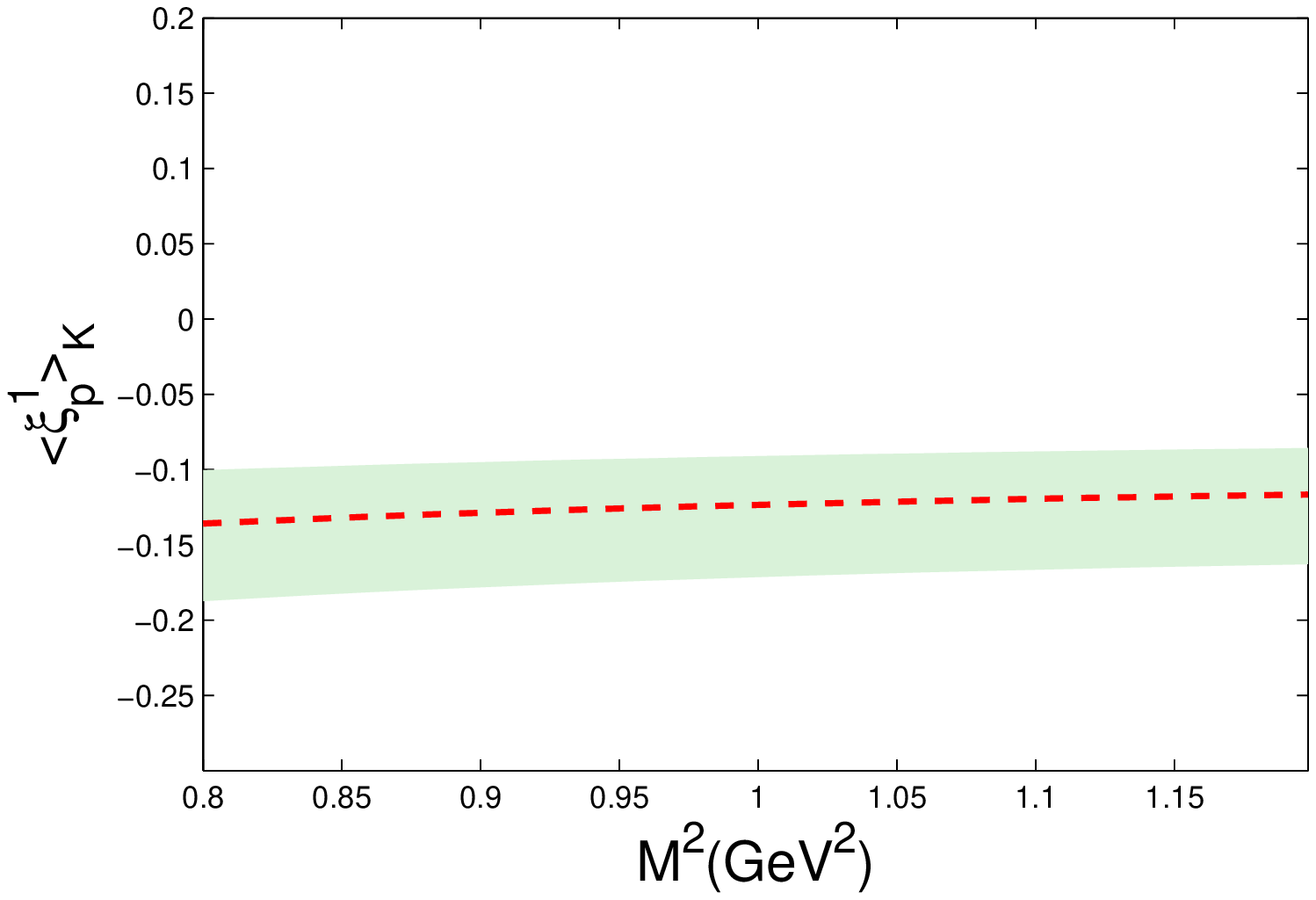}
\includegraphics[width=0.24\textwidth]{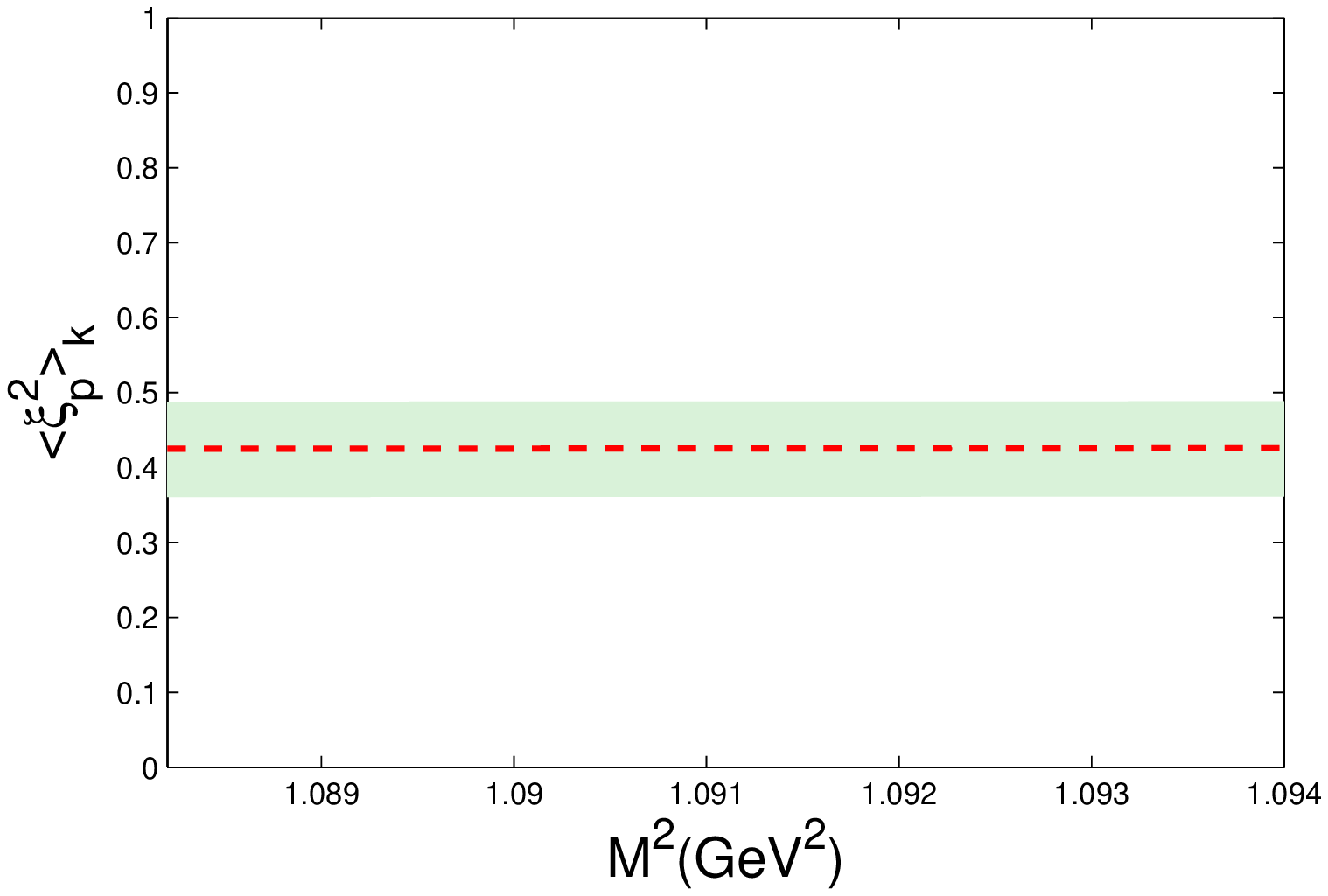}
\includegraphics[width=0.24\textwidth]{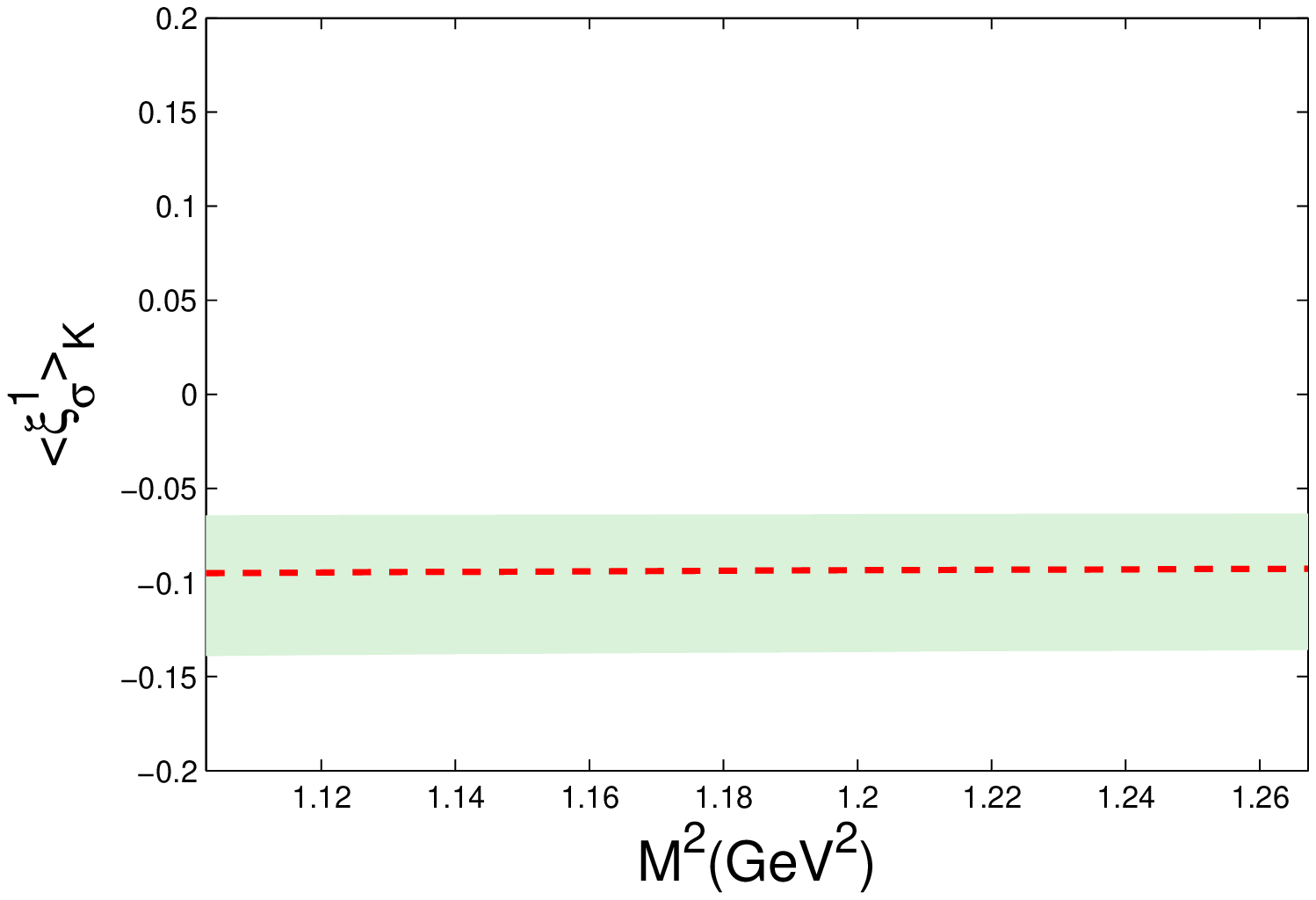}
\includegraphics[width=0.24\textwidth]{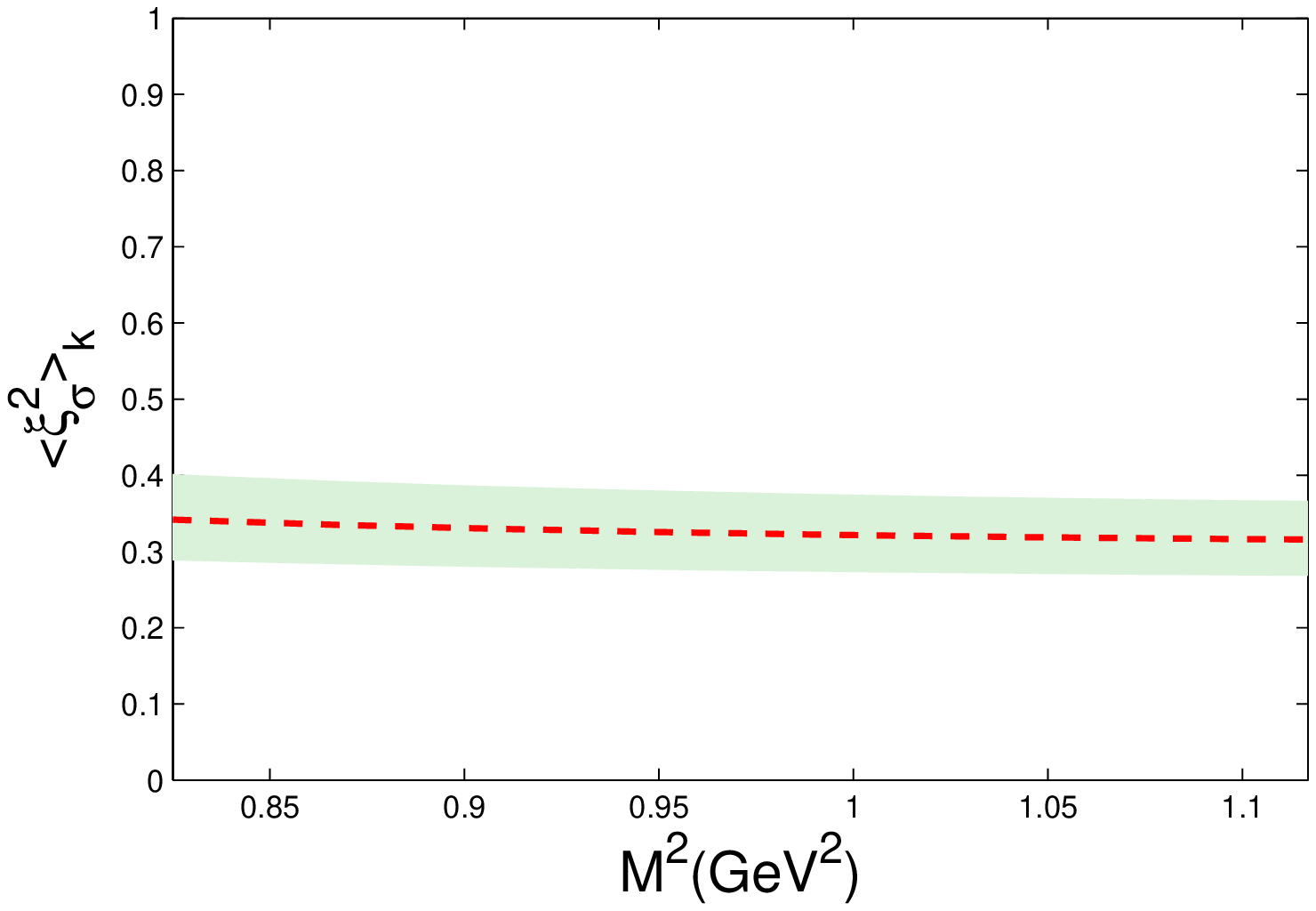}
\caption{The first and second moments of kaon twist-3 DAs versus the Borel parameter $M^2$, where the shaded bands are the uncertainties caused by varying all the input parameters within their reasonable regions.}
\label{xik}
\end{figure}
\end{center}
\end{widetext}

Second, we calculate the first two moments of kaon twist-3 DAs. The Borel window for the second moment of $\left<\xi^2_{p,\sigma} \right>_K$ is determined by setting the continuum contribution to be less than $35\%$ and the dimension-six condensate contribution to be less than $3\%$. The Borel window for the first moment of $\left<\xi^1_{p,\sigma} \right>_K$ is determined by setting the continuum contribution to be less than $5\%$ for $\left<\xi^1_\sigma\right>_K$ and to be less than $3\%$ for $\left<\xi^1_p\right>_K$; while the dimension-six condensate contribution is set to be less than $1\%$ for $\left<\xi^1_{p}\right>_K$ and less than $3\%$ for $\left<\xi^1_{\sigma}\right>_K$. The results together with the corresponding uncertainties are presented in Tab.\ref{tsrK} and Tab.\ref{uncertK}. To show the uncertainties more clearly, we draw the first two moments of the kaon twist-3 DAs versus $M^2$ in Fig(\ref{xik}), where the shaded bands are the uncertainties caused by varying all the input parameters within their reasonable regions. By adding these uncertainties in quadrature, and with the help of the relation between different moments (\ref{gegen_moment}) and the scale running relation (\ref{rge}), we can obtain the corresponding Gegenbauler moments at the scale $\mu_f = 1$ GeV:
\begin{eqnarray}
a^1_{K,p}(1 {\rm GeV}) &=& -0.376^{+0.103}_{-0.148}, \\
a^2_{K,p}(1 {\rm GeV}) &=& 0.701^{+0.481}_{-0.491}, \\
a^1_{K,\sigma}(1 {\rm GeV}) &=& -0.160^{+0.051}_{-0.074}, \\
a^2_{K,\sigma}(1 {\rm GeV}) &=& 0.369^{+0.163}_{-0.149}.
\end{eqnarray}

\begin{widetext}
\begin{center}
\begin{table}
\caption{Wavefunction parameters for $\psi^K_p(x,\textbf{k}_\perp)$ with typical DA moments at $\mu_f = 1$ GeV.}
\begin{tabular}{|c| c c c | c c c | c c c|}
\hline\hline
$a^1_{K,p}$ & & $\ -0.524\quad$& & & $\ -0.376\quad$ & & & $\ -0.273\quad$ & \\
$a^2_{K,p}$ & $\  0.210 \quad$ & $\  0.701 \quad$ & $\  1.182 \quad$ & $\  0.210 \quad$ & $\  0.701 \quad$ & $\  1.182 \quad$ & $\  0.210 \quad$ & $\  0.701 \quad$ & $\  1.182 \quad$ \\
\hline
$A^K_p(\textrm{GeV}^{-2})$ & $163.047$ & $139.472$ & $126.284$ & $168.526$ & $143.459$ & $129.680$ & $172.755$ & $146.721$ & $132.586$ \\
$B^K_p$                    & $-0.249$ & $-0.188$ & $-0.137$ & $-0.063$ & $-0.019$ & $0.021 $ & $0.064 $ & $0.096 $ & $0.129 $ \\
$C^K_p$                    & $1.133 $ & $1.559 $ & $1.968 $ & $1.168 $ & $1.583 $ & $1.986 $ & $1.194 $ & $1.601 $ & $1.999 $ \\
$\beta^K_p(\textrm{GeV})$  & $0.506 $ & $0.549 $ & $0.581 $ & $0.503 $ & $0.546 $ & $0.578 $ & $0.501 $ & $0.544 $ & $0.575 $ \\
\hline\hline
\end{tabular}
\label{psipk1}
\end{table}

\begin{table}
\caption{Wavefunction parameters for $\psi^K_\sigma(x,\textbf{k}_\perp)$ with typical DA moments at $\mu_f = 1$ GeV.}
\begin{tabular}{|c| c c c | c c c | c c c|}
\hline\hline
$a^1_{K,\sigma}$ & & $\ -0.234\quad$& & & $\ -0.160\quad$ & & & $\ -0.109\quad$ & \\
$a^2_{K,\sigma}$ & $\  0.220 \quad$ & $\  0.369 \quad$ & $\  0.532 \quad$ & $\  0.220 \quad$ & $\  0.369 \quad$ & $\  0.532 \quad$ & $\  0.220 \quad$ & $\  0.369 \quad$ & $\  0.532 \quad$ \\
\hline
$A^K_\sigma(\textrm{GeV}^{-2})$ & $194.004$ & $150.888$ & $116.789$ & $199.642$ & $154.708$ & $119.355$ & $203.549$ & $157.461$ & $121.467$ \\
$B^K_\sigma$                    & $-0.080$ & $-0.062$ & $-0.049$ & $-0.007$ & $0.010 $ & $0.026 $ & $0.043 $ & $0.059 $ & $0.075 $ \\
$C^K_\sigma$                    & $0.114 $ & $0.225 $ & $0.349 $ & $0.129 $ & $0.237 $ & $0.360 $ & $0.138 $ & $0.246 $ & $0.367 $ \\
$\beta^K_\sigma(\textrm{GeV})$  & $0.442 $ & $0.479 $ & $0.519 $ & $0.440 $ & $0.477 $ & $0.516 $ & $0.438 $ & $0.475 $ & $0.514 $ \\
\hline\hline
\end{tabular}
\label{psisik1}
\end{table}
\end{center}
\end{widetext}

\begin{center}
\begin{figure}
\includegraphics[width=0.35\textwidth]{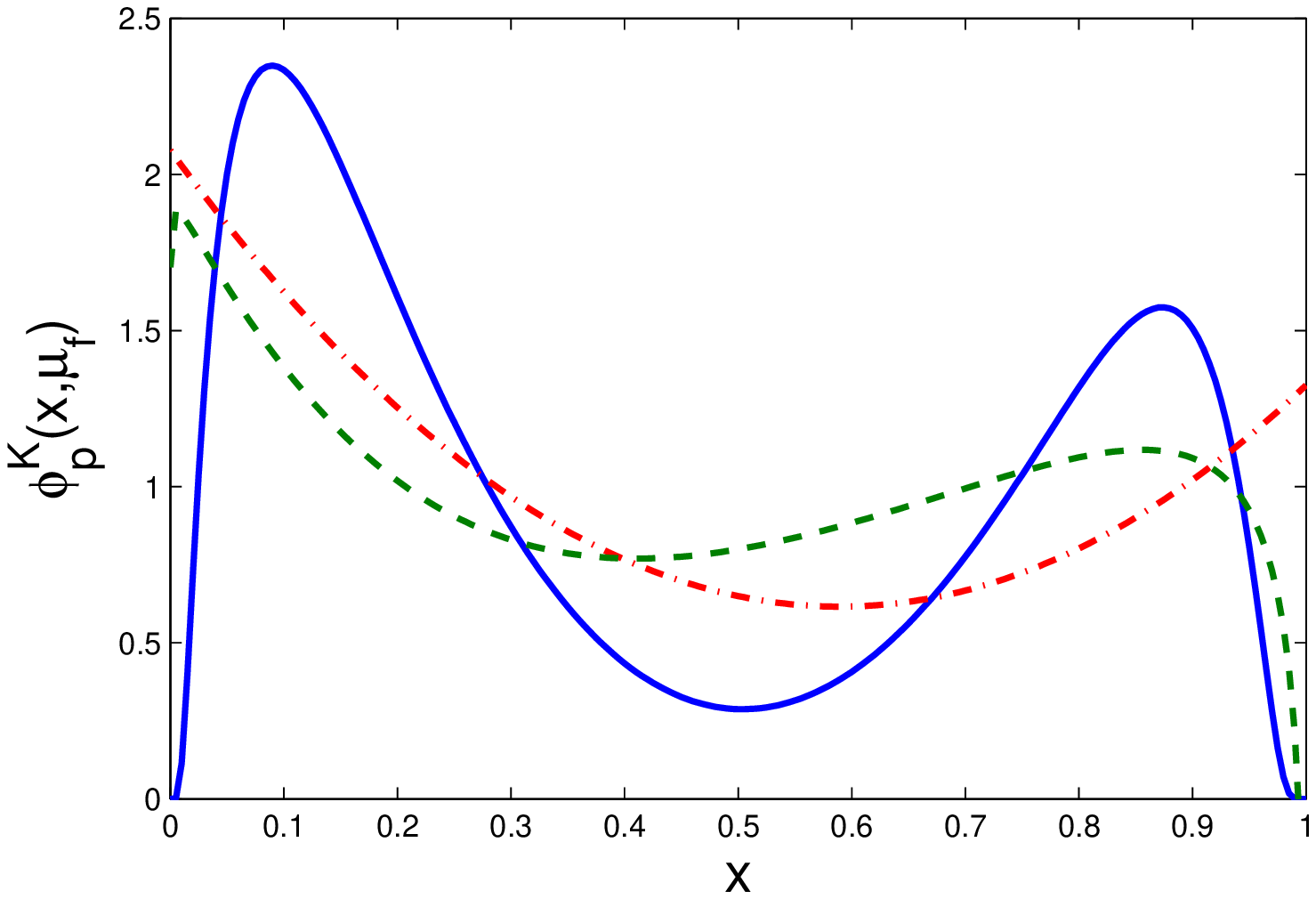}
\includegraphics[width=0.35\textwidth]{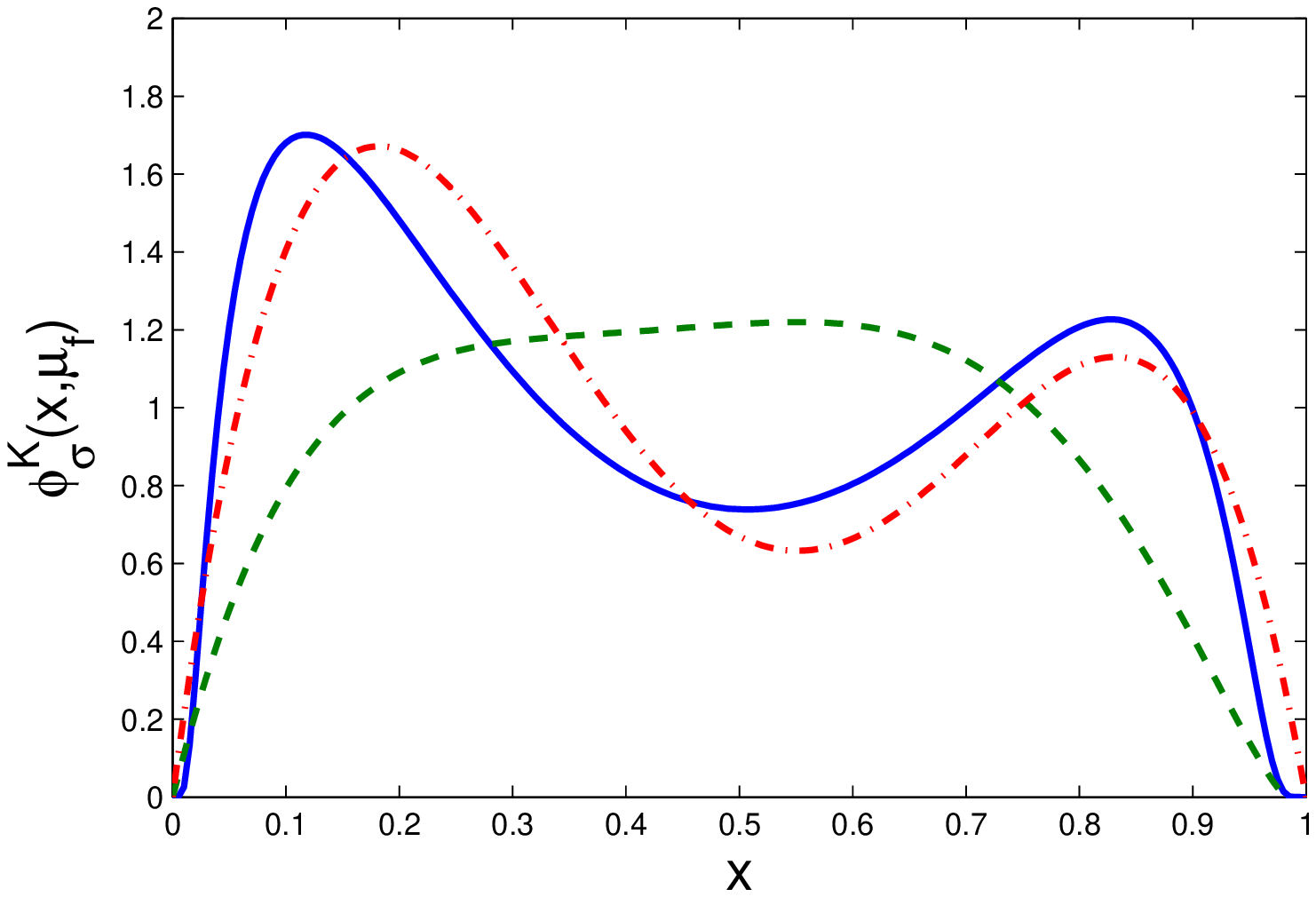}
\caption{Kaon twist-3 DAs $\phi^K_p(x,\mu_f)$ (Left) and $\phi^K_\sigma(x,\mu_f)$ (Right), where the solid and the dash-dot lines are for our DAs defined by Eq.(\ref{phips}) and Eqs.(\ref{phi_g_m0},\ref{phi_g_m1}) with $\mu_f = 1$ GeV respectively. As a comparison, the DA of Ref.(\cite{T3_K}) at $\mu_f = 1$ GeV is shown by a dashed line. }
\label{fphipk}
\end{figure}
\end{center}

Based on the above moments and the formulas presented in Sec.II, we can obtain the kaon twist-3 wave function parameters, which are collected in Tabs.(\ref{psipk1},\ref{psisik1}). Here Tabs.(\ref{psipk1},\ref{psisik1}) correspond to the factorization scale $\mu_f = 1$ GeV. With the help of Eq.(\ref{phips}), we can obtain the kaon twist-3 DAs $\phi^K_p$ and $\phi^K_\sigma$, which are presented in Fig.(\ref{fphipk}). Our $\phi^K_p$ (Left diagram) or $\phi^K_\sigma$ (Right diagram) are drawn by the solid lines which are defined by Eq.(\ref{phips}), and the dash-dot lines are for Eqs.(\ref{phi_g_m0},\ref{phi_g_m1}) with $\mu_f = 1$ GeV, respectively. As a comparison, we also give the DAs of Ref.\cite{T3_K} under $\mu_f = 1$ GeV, which are drawn by the dashed lines. Here, in doing the comparison, we need to replace $\phi^K_{p,\sigma}(x)$ in Ref.\cite{T3_K} to be $\phi^K_{p,\sigma}(1-x)$, because in Ref.\cite{T3_K} $x$ stands for the momentum fraction of $s$-quark; while in the present paper, $x$ is taken as that of $u$ (or $d$) quark. These two figures indicate that the kaon twist-3 DAs, especially $\phi^K_p$, have a better end-point behavior. Such a BHL-improved behavior shall be helpful to obtain a reasonable result for kaon related processes, such as the kaon electromagnetic form factor and the $B \to K$ transition form factor with $k_T$ factorization approach or LCSR. Some previous calculations can be found in Refs.\cite{BK_ff,KE,bhlc}.

\begin{widetext}
\begin{center}
\begin{table}
\caption{Contributions of the higher order quark mass terms to various vacuum condensate parts for the moments of $\phi^K_{p,\sigma}$. For estimation of the contribution of higher order quark mass terms to every vacuum condensate, the percentage of which is obtained by calculating the ratio of higher order mass terms ($m^n$, $n\geq 2$) before each vacuum condensate with those of ($m^n$, $n\leq 1$). }
\begin{tabular}{|c||c|c||c|c|}
\hline \hline
 & $\left<\xi^1_p\right>_K$ & $\left<\xi^2_p\right>_K$ & $\left<\xi^1_\sigma\right>_K$ & $\left<\xi^2_\sigma\right>_K$ \\  \hline\hline
$\left<\alpha_sG^2\right>$              & $4.3\%\sim2.8\%$ & $2.3\%\sim2.2\%$  & $-$ & $-$ \\  \hline
$m_q\left<\bar{q}q\right>$              & $-0.7\%\sim-0.4\%$   & $17.6\%\sim17.5\%$ & $1.2\%\sim1.0\%$  & $2.8\%\sim1.9\%$ \\  \hline
$m_q\left<g_s\bar{q}\sigma TGq\right>$  & $7.2\%\sim4.4\%$    & $-8.0\%\sim-7.9\%$       & $11.7\%\sim10.0\%$  & $-0.6\%\sim-0.4\%$ \\  \hline
$\left<g_s\bar{q}q\right>^2$            & $-16.5\%\sim-9.1\%$   & $-0.5\%$  & $-1.6\%\sim-1.3\%$  & $1.0\%\sim 0.7\%$ \\  \hline \hline
\end{tabular}
\label{mck}
\end{table}
\end{center}
\end{widetext}

As has been argued in the Introduction in order to provide a more sound estimation on the $SU_f(3)$-breaking effect in the $K$-meson involved processes, we need to use the much more complex Eq.(\ref{prop of s3}) other than Eq.(\ref{lqp}) as the quark propagator. To show this point clearly, we show in Tab.\ref{mck} how the higher order mass terms contribute to the corresponding vacuum condensates for the first two moments of $\phi^K_{p,\sigma}$. For estimation of the contribution of higher order quark mass terms to every vacuum condensate, e.g. $\left<\alpha_sG^2\right>$, $m_q\left<\bar{q}q\right>$, $m_q\left<g_s\bar{q}\sigma TGq\right>$ or $\left<g_s\bar{q}q\right>^2$, the percentage of which is obtained by calculating the ratio of higher order mass terms ($m^n$, $n\geq 2$) before each vacuum condensate with those of ($m^n$, $n\leq 1$). Tab.\ref{mck} indicates that because the $s$-quark mass is not small, it shall lead to sizable contributions. For example, its contribution to $m_q\left<\bar{q}q\right>$ for $\left<\xi^{2}_{p}\right>_K$ can be up to $17\%$.

\section{summary}

The background field approach provides a systematic description for the vacuum condensates from the viewpoint of field theory and it provides a convenient way to derive the QCD sum rules. We have made an investigation over the kaon twist-3 DAs $\phi_{p,\sigma}^K$ within this approach. Furthermore, the $SU_f(3)$-breaking effects are studied in detail under a more systematical way, especially the quark propagator (\ref{prop of s3}) that keeps the mass terms consistently is adopted. As have been shown by Tab.\ref{mck}, higher-order mass terms can indeed provide sizable contributions to the kaon DA moments. For example, its contribution to $m_q\left<\bar{q}q\right>$ for $\left<\xi^{2}_{p}\right>_K$ can be up to $17\%$. So to obtain a sound estimation for the $SU_f(3)$-breaking effect, we need to take these higher-order mass terms into consideration. Moreover, such a propagator shall also be helpful for deriving information on the meson or baryon with heavy quarks. Some more works on its application to the heavy meson/baryon properties are in progress.

As for the kaon twist-3 DAs $\phi^K_{p}$ and $\phi^K_{\sigma}$, we have studied their normalization parameters and moments within the QCD sum rules under the background field approach. For its normalization parameters, we obtain $\mu^p_K |_{1 \rm GeV} = 1.188^{+0.039}_{-0.043}$ GeV and $\mu_K^\sigma |_{1\rm GeV} = 1.021^{+0.036}_{-0.055}$ GeV. As for the moments of $\phi^K_{p,\sigma}$, around $\mu_f \simeq 1 \rm GeV$, we obtain $\left<\xi^1_p\right>_K = -0.126^{+0.034}_{-0.050}$, $\left<\xi^2_p\right>_K = 0.425^{+0.063}_{-0.064}$, $\left<\xi^1_\sigma\right>_K = -0.094^{+0.030}_{-0.044}$ and $\left<\xi^2_\sigma\right>_K = 0.329^{+0.057}_{-0.052}$. Basing on these moments, we further calculate the Gegenbauler moments, and  establish a model for kaon twist-3 wavefunctions $\Phi^K_{p,\sigma}$ with the help of BHL prescription, which have a better endpoint behavior and shall be helpful for estimating the kaon involved inclusive or exclusive processes.

\begin{widetext}
\begin{center}
\begin{table}
\caption{The normalization parameters and the first two moments of $\phi^\pi_{p,\sigma}$ and their corresponding Borel windows. }
\begin{tabular}{|c|c|c|c||c|c|}
\hline \hline
$M^2(\rm GeV^2)$ & $[0.799,0.976]$ & $M^2(\rm GeV^2)$                & $[0.895,1.066]$   & $M^2(\rm GeV^2)$                & $[0.833,1.109]$\\
$(\mu_\pi^{p})^2(\rm GeV^2)$ & $1.068 - 1.280$ & $\left<\xi^2_p\right>_\pi$      & $0.539 \pm 0.012$ & $\left<\xi^4_p\right>_\pi$      & $0.580 \pm 0.010$\\
\hline\hline
$M^2(\rm GeV^2)$ & $[0.974,1.052]$ & $M^2(\rm GeV^2)$                & $[0.831,1.163]$   & $M^2(\rm GeV^2)$                & $[1.095,1.340]$\\
$(\mu_\pi^p \mu_\pi^\sigma)(\rm GeV^2)$ & $1.229 - 1.316$ & $\left<\xi^2_\sigma\right>_\pi$ & $0.313 \pm 0.012$ & $\left<\xi^4_\sigma\right>_\pi$ & $0.248 \pm 0.004$\\
\hline \hline
\end{tabular}
\label{tsrPi}
\end{table}
\end{center}
\end{widetext}

As a final remark, by setting the current quark mass $m_s=0$, we can obtain the sum rules for the pion distribution amplitudes $\phi^{\pi}_{p,\sigma}$, whose normalization parameters and the first two non-zero moments together with their corresponding Borel windows are presented in Tab.\ref{tsrPi}. By adding all the uncertainties in quadrature, we obtain $\mu^p_\pi |_{1 \rm GeV} = 1.104^{+0.046}_{-0.050}$ GeV and $\mu_\pi^\sigma |_{1 \rm GeV} = 1.149^{+0.033}_{-0.034}$ GeV. If taking the $\alpha_s$ correction into consideration, which increases the leading-order results by $15\%-20\%$ \cite{nlomu}, we shall obtain
$\mu^p_\pi |_{1 \rm GeV} \in [1.20,1.38]$ GeV and $\mu_\pi^\sigma |_{1 \rm GeV} \in [1.28,1.42]$ GeV.

\section*{Acknowledgements}

This work was supported in part by the Fundamental Research Funds for the Central Universities under Grant No.CDJXS1102209 and the Program for New Century Excellent Talents in University under Grant No. NCET-10-0882, and by Natural Science Foundation of China under Grant No.11075225.

\appendix

\section{Details for the formulas of $\left<0\left| \bar{\psi}^a_\alpha (x) \psi^b_\beta (y) \right|0\right>$ and $\left<0\left| \bar{\psi}^a_\alpha (x) \psi^b_\beta (y) G^A_{\mu\nu} \right|0\right>$}

Under the background field approach, $\bar{\psi}(x)$ can be expanded around $x=0$ \cite{pro_func},
\begin{eqnarray}
\bar{\psi}(x) &=& \bar{\psi}(0) + \bar{\psi}(0) \overleftarrow{D}^\alpha x_\alpha + \frac{1}{2} \bar{\psi}(0) \overleftarrow{D}^\alpha \overleftarrow{D}^\beta x_\alpha x_\beta + \nonumber\\
&& \frac{1}{3!} \bar{\psi}(0) \overleftarrow{D}^\alpha \overleftarrow{D}^\beta \overleftarrow{D}^\gamma x_\alpha x_\beta x_\gamma + \cdots . \nonumber
\end{eqnarray}
Then, the matrix element $\left<0\left| \bar{\psi}^a_\alpha(x) \psi^b_\beta(0) \right|0\right>$ can be expanded around $x=0$ as
\begin{eqnarray}
&& \left<0\left| \bar{\psi}^a_\alpha(x) \psi^b_\beta(0) \right|0\right> \nonumber\\
&= & \left<0\left| \bar{\psi}^a_\alpha(0) \psi^b_\beta(0) \right|0\right> + x_\mu \left<0\left| \bar{\psi}^a_\alpha(0) \overleftarrow{D}^\mu \psi^b_\beta(0) \right|0\right>  \nonumber\\
&& + \frac{1}{2!} x_\mu x_\nu \left<0\left| \bar{\psi}^a_\alpha(0) \overleftarrow{D}^\mu \overleftarrow{D}^\nu \psi^b_\beta(0) \right|0\right> \nonumber\\
&& + \frac{1}{3!} x_\mu x_\nu x_\rho \left<0\left| \bar{\psi}^a_\alpha(0) \overleftarrow{D}^\mu \overleftarrow{D}^\nu \overleftarrow{D}^\rho \psi^b_\beta(0) \right|0\right>.
\label{qqx}
\end{eqnarray}
The results for the first and the second terms are well known,
\begin{displaymath}
\left<0\left| \bar{\psi}^a_\alpha(0) \psi^b_\beta(0) \right|0\right> = \frac{1}{12} \left<\bar{\psi}\psi\right> \delta^{ab} g_{\alpha\beta}
\end{displaymath}
and
\begin{displaymath}
\left<0\left| \bar{\psi}^a_\alpha(0) \overleftarrow{D}^\mu \psi^b_\beta(0) \right|0\right> = \frac{im}{48} \left<\bar{\psi}\psi\right> \delta^{ab} (\gamma^\mu)_{\beta\alpha} \ .
\end{displaymath}
As for the third term $\left<0\left| \bar{\psi}^a_\alpha(0) \overleftarrow{D}^\mu \overleftarrow{D}^\nu \psi^b_\beta(0) \right|0\right>$, basing on its color and Dirac-gamma structures, it can be rewritten as
\begin{displaymath}
\left<0\left| \bar{\psi}^a_\alpha(0) \overleftarrow{D}^\mu \overleftarrow{D}^\nu \psi^b_\beta(0) \right|0\right> = C g^{\mu\nu} g_{\beta\alpha}\delta^{ba}+ D(\sigma^{\mu\nu})_{\beta\alpha} \delta^{ba} .
\end{displaymath}
Utilizing the equation of motion of the background quark field and the equation $[ \overleftarrow{D}^\mu , \overleftarrow{D}^\nu ] = -ig_sT^A G^{A\mu\nu}$, we obtain $C = -\frac{m^2}{48} \left<\bar{\psi}\psi\right> + \frac{1}{96} \left< g_s \bar{\psi} \sigma T G \psi \right>$ and $D = -\frac{i}{288} \left< g_s \bar{\psi} \sigma T G \psi \right>$, where $\left< g_s \bar{\psi} \sigma TG \psi \right>$ is the abbreviation of $\left< g_s \bar{\psi} \sigma_{\mu\nu} T^A G^{A\mu\nu} \psi \right>$. The fourth term $\left<0\left| \bar{\psi}^a_\alpha(0) \overleftarrow{D}^\mu \overleftarrow{D}^\nu \overleftarrow{D}^\rho \psi^b_\beta(0) \right|0\right>$ can be treated similarly,
\begin{eqnarray}
&&\left<0\left| \bar{\psi}^a_\alpha(0) \overleftarrow{D}^\mu \overleftarrow{D}^\nu \overleftarrow{D}^\rho \psi^b_\beta(0) \right|0\right>  \nonumber\\
&&= \delta^{ba} ( E \gamma^\mu g^{\nu\rho} + F \gamma^\nu g^{\nu\rho} + G \gamma^\rho g^{\mu\nu} )_{\beta\alpha} \nonumber
\end{eqnarray}
with $E = G = -\frac{im^3}{96 \times 3} \left<\bar{\psi}\psi\right> + \frac{im}{96 \times 9} \left< g_s \bar{\psi} \sigma T G \psi \right>$ and  $F = -\frac{im^3}{96 \times 3} \left<\bar{\psi}\psi\right> + \frac{4im}{96 \times 9} \left< g_s \bar{\psi} \sigma T G \psi \right> $. Taking use of translation invariance of matrix element, we finally obtain
\begin{widetext}
\begin{eqnarray}
\left<0\left| \bar{\psi}^a_\alpha(x) \psi^b_\beta(y) \right|0\right> &=& \delta^{ba} \left\{ \left<\bar{\psi}\psi\right> \left[ \frac{1}{12} g_{\beta\alpha} + \frac{im}{48} (\not\! x - \not\! y)_{\beta\alpha} - \frac{m^2}{96} (x-y)^2 g_{\beta\alpha}- \frac{im^3}{96 \times 6} (x-y)^2 (\not\! x - \not\! y)_{\beta\alpha} \right] \right. \nonumber\\
&& \left. + \left < g_s \bar{\psi} \sigma TG \psi \right> \left[ \frac{1}{96 \times 2} (x-y)^2 g_{\beta\alpha} + \frac{im}{96 \times 9} (x-y)^2 (\not\! x - \not\! y)_{\beta\alpha} \right] \right\} .
\label{qqxy}
\end{eqnarray}
\end{widetext}

The matrix element $\left<0\left| \bar{\psi}^a_\alpha (x) \psi^b_\beta (y) G^A_{\mu\nu} \right|0\right>$ can also be expanded it around $x=0$,
\begin{eqnarray}
&&\left<0\left| \bar{\psi}^a_\alpha (x) \psi^b_\beta (y) G^A_{\mu\nu} \right|0\right> =\nonumber\\
&& \left<0\left| \bar{\psi}^a_\alpha (0) \psi^b_\beta (0) G^A_{\mu\nu} \right|0\right> + x_\rho \left<0\left| \bar{\psi}^a_\alpha (0) \overleftarrow{D}^\rho G^A_{\mu\nu} \psi^b_\beta(0)\right|0\right>  \nonumber\\
&& + y_\rho \left<0\left| \bar{\psi}^a_\alpha (0) G^A_{\mu\nu} \overrightarrow{D}^\rho \psi^b_\beta (0) \right|0\right>.
\end{eqnarray}
Obviously,
\begin{displaymath}
\left<0\left| \bar{\psi}^a_\alpha (0) \psi^b_\beta (0) G^A_{\mu\nu} \right|0\right> = \frac{1}{192} \left< \bar{\psi} \sigma TG \psi \right> (\sigma_{\mu\nu})_{\beta\alpha} (T^A)^{ba}.
\end{displaymath}
Utilizing the equation $[T^A G^A_{\mu\nu} , \overleftarrow{D}^\rho] = T^A G^{A\ \ \rho}_{\mu\nu ;}$, one can derive
\begin{eqnarray}
&&\left<0\left| \bar{\psi}^a_\alpha (0) \overleftarrow{D}^\rho G^A_{\mu\nu} \psi^b_\beta (0) \right|0\right> = \nonumber\\
&& -\left<0\left| \bar{\psi}^a_\alpha (0) G^A_{\mu\nu} \overrightarrow{D}^\rho \psi^b_\beta (0) \right|0\right>  - \left<0\left| \bar{\psi}^a_\alpha (0) \psi^b_\beta (0) G^{A\ \ \rho}_{\mu\nu;} \right|0\right> .\nonumber
\end{eqnarray}
Using the equation of motion of the background quark field together with the following equation \cite{vac_con2}:
\begin{eqnarray}
&&\left<0\left| \bar{\psi}^a_\alpha (0) \psi^b_\beta (0) G^{A\ \ \rho}_{\mu\nu;} \right|0\right> = \nonumber\\
&&\quad\quad\quad \frac{1}{432} g_s \left<\bar{\psi}\psi\right>^2 (g_{\rho\mu} \gamma_\nu - g_{\rho\nu} \gamma_\mu)_{\beta\alpha} (T^A)^{ba}, \nonumber
\end{eqnarray}
we obtain,
\begin{eqnarray}
&&\left<0\left| \bar{\psi}^a_\alpha (0) \overleftarrow{D}^\rho G^A_{\mu\nu} \psi^b_\beta (0) \right|0\right> = \nonumber\\
&& -\frac{1}{864} g_s \left<\bar{\psi}\psi\right>^2 (g_{\rho\mu} \gamma_\nu - g_{\rho\nu} \gamma_\mu)_{\beta\alpha} (T^A)^{ba} + \nonumber\\
&& \left[ \frac{im}{384} \left< \bar{\psi} \sigma TG \psi \right>
+ \frac{i}{864} g_s \left<\bar{\psi}\psi\right>^2 \right] (\epsilon_{\rho\mu\nu\sigma} \gamma_5 \gamma^\sigma)_{\beta\alpha} (T^A)^{ba} \nonumber
\end{eqnarray}
and
\begin{eqnarray}
&&\left<0\left| \bar{\psi}^a_\alpha (0) G^A_{\mu\nu} \overrightarrow{D}^\rho \psi^b_\beta (0) \right|0\right> =\nonumber\\
&& -\frac{1}{864} g_s \left<\bar{\psi}\psi\right>^2 (g_{\rho\mu} \gamma_\nu - g_{\rho\nu} \gamma_\mu)_{\beta\alpha} (T^A)^{ba}-  \nonumber\\
&& \left[ \frac{im}{384} \left< \bar{\psi} \sigma TG \psi \right>
+ \frac{i}{864} g_s \left<\bar{\psi}\psi\right>^2 \right] (\epsilon_{\rho\mu\nu\sigma} \gamma_5 \gamma^\sigma)_{\beta\alpha} (T^A)^{ba}. \nonumber
\end{eqnarray}
Then, we finally have
\begin{widetext}
\begin{eqnarray}
\left<0\left| \bar{\psi}^a_\alpha (x) \psi^b_\beta (y) G^A_{\mu\nu} \right|0\right> &=& \frac{1}{192} \left< \bar{\psi} \sigma TG \psi \right> (\sigma_{\mu\nu})_{\beta\alpha} (T^A)^{ba} + \left\{ -\frac{1}{864} g_s \left<\bar{\psi}\psi\right>^2 (g_{\rho\mu} \gamma_\nu - g_{\rho\nu} \gamma_\mu)_{\beta\alpha} (x+y)^\rho
\right. \nonumber\\
&& \left. + i(x-y)^\rho \left[ \frac{m}{384} \left< \bar{\psi} \sigma TG \psi \right> + \frac{1}{864} g_s \left<\bar{\psi}\psi\right>^2 \right] (\epsilon_{\rho\mu\nu\sigma} \gamma_5 \gamma^\sigma)_{\beta\alpha} \right\} (T^A)^{ba}.
\label{qqgxy}
\end{eqnarray}
\end{widetext}
It is found that Eqs.(\ref{qqxy},\ref{qqgxy}) agree with those of Ref.\cite{vac_con_LJW} (Eqs.(22,29) there), except that for $\left<0\left| \bar{\psi}^a_\alpha(x) \psi^b_\beta(y) \right|0\right>$ there is no dimension-six term and the coefficient before $m \left< g_s \bar{\psi} \sigma TG \psi \right>$ should be $\frac{i}{96 \times 9}$ other than $\frac{i}{96 \times 12}$, and for $\left<0\left| \bar{\psi}^a_\alpha (x) \psi^b_\beta (y) G^A_{\mu\nu} \right|0\right>$ the last term should be $(x-y)^\rho$ other than $(y-x)^\rho$.


\begin{thebibliography}{99}

\bibitem{LCSR1} V. L. Chernyak and I. R. Zhitnitsky, Nucl.Phys. B{\bf 345} (1990) 137.

\bibitem{LCSR2} I. I. Balitsky, V. M. Braun, and A. V. Kolesnichenko, Nucl.Phys. B{\bf 312} (1989) 509.

\bibitem{excu1} G. P. Lepage and S. J. Brodsky, Phys.Rev. D{\bf 22} (1980) 2157.

\bibitem{excu2} S. J. Brodsky and G. P. Lepage, Phys.Rev. D{\bf 24} (1981) 1808.

\bibitem{excu3} V.L. Chernyak and A.R. Zhitnitsky, Phys.Rept. {\bf 112} (1984) 173.

\bibitem{T3_85} A. R. Zhitnitsky, I. R. Zhitnitsky, and V. L. Chernyak, Yad.Fiz. {\bf 41} (1985) 445.

\bibitem{T3_90} V. M. Braun and I. E. Filyanov, Z.Phys. C{\bf 48} (1990) 239.

\bibitem{T3_98_PBall} P. Ball, V. M. Braun, Y. Koike, and K. Tanaka, Nucl. Phys. B{\bf 529} (1998) 323.

\bibitem{T3_04_HT} T. Huang, X. H. Wu, and M. Z. Zhou, Phys.Rev. D{\bf 70} (2004) 014013.

\bibitem{PI_T3_MODEL} T. Huang and X. G. Wu, Phys. Rev. D{\bf 70} (2004) 093013.

\bibitem{T3_11_ZHONG} T. Zhong, {\it et al.}, Phys.Rev. D{\bf 83} (2011) 036002.

\bibitem{T3_99_PBall} P. Ball, JHEP {\bf 9901} (1999) 010.

\bibitem{T3_K} P. Ball, V. M. Braun and A. Lenz, JHEP {\bf 0605} (2006) 004.

\bibitem{T3_05_HT} T. Huang, M. Z. Zhou, and X. H. Wu, Eur.Phys.J. C{\bf 42}  (2005) 271.

\bibitem{twist31} Seung-il Nam and H.C. Kim, Phys.Rev. D{\bf 74}  (2006) 096007.

\bibitem{scale} P. Ball and R. Zwicky, Phys.Rev. D{\bf 71} (2005) 014015.

\bibitem{BK_ff} X.G. Wu, T. Huang, and Z.Y. Fang, Eur.Phys.J. C{\bf 52} (2007) 561.

\bibitem{KE} X.G. Wu and T. Huang, JHEP {\bf 0804} (2008) 043.

\bibitem{BG1} V. A. Novikov, M. A. Shifman, A. I. Vainshtein, and V. I. Zakharov, Fortschr. Phys. {\bf 32} (1984) 585 .

\bibitem{BG2} W. Hubschmid and S. Mallik, Nucl. Phys. B{\bf 207} (1982) 29;

\bibitem{BG3} J. Govaerts, F. de Viron, D. Gusbin, and J. Weyers, Phys.Lett. B{\bf 128} (1983) 262; Nucl. Phys. B{\bf 248} (1984) 1;

\bibitem{BG4} J. Ambjorn and R.J. Hughes, Annals Phys. {\bf 145} (1983) 340; Nucl.Phys. B{\bf 217} (1983) 336.

\bibitem{BG_HT} T. Huang and Z. Huang, Phys.Rev. D{\bf 39} (1989) 1213.

\bibitem{pro_func} T. Huang, X. N. Wang, and X. D. Xiang, Phys. Rev. D{\bf 35}  (1987) 1013.

\bibitem{svz} M.A. Shifman, A.I. Vainshtein, and V.I. Zakharov, Nucl.Phys. B{\bf 147} (1979) 385.

\bibitem{pro1} L. J. Reinders, H. Rubinstein, and S. Yazaki, Phys.Rept. {\bf 127} (1985) 1.

\bibitem{fixed1} M.A. Shiftman, Nucl.Phys. B{\bf 173}  (1980) 13.

\bibitem{fixed2} M.S. Dubovikov and A.V. Smilga, Nucl.Phys. B{\bf 185} (1981) 109.

\bibitem{ddpi} V.M. Belyaev, V.M. Braun, A. Khodjamirian, and R. Ruckl, Phys.Rev. D{\bf 51} (1995) 6177.

\bibitem{bhld} T. Huang and X.G. Wu, Int.J.Mod.Phys. A{\bf 22}  (2007) 3065.

\bibitem{bhla} T. Huang, B.Q. Ma, and Q.X. Shen, Phys.Rev. D{\bf 49} (1994) 1490.

\bibitem{bhlb}  B.W. Xiao and B.Q. Ma, Phys.Rev. D{\bf 71}  (2005) 014034; B.W. Xiao, X. Qian, and B.Q. Ma, Eur.Phys.J. A{\bf 15} (2002) 523.

\bibitem{bhlc} T. Huang, X.G. Wu, and X.H. Wu, Phys.Rev. D{\bf 70} (2004) 053007; X.G. Wu, T. Huang, and Z. Y. Fang, Phys. Rev. D{\bf 77}  (2008) 074001; X. G. Wu and T. Huang, Phys. Rev. D{\bf 79} (2009) 034013.

\bibitem{BHL} S. J. Brodsky, T. Huang, and G. P. Lepage, in Proceedings of the Banff Summer Institute, Banff, Alberta, 1981, edited by A. Z. Capri and A. N. Kamal (Plenum, New York, 1983), p. 143; G. P. Lepage, S. J. Brodsky, T. Huang, and P. B. Mackenize, ibid., p. 83; T. Huang, in Proceedings of XXth International Conference on High Energy Physics, Madison, Wisconsin, 1980, edited by L. Durand and L. G. Pondrom, AIP Conf. Proc. No. 69 (AIP, New York, 1981), p. 1000.

\bibitem{transverse_momentum} X.H. Guo and T. Huang, Phys. Rev. D{\bf 43} (1991) 2931.

\bibitem{PDG} K. Nakamura \textit{et al}. (Particle Data Group), J. Phys. G{\bf 37} (2010) 075021 .

\bibitem{VC80} A. Zalewska and K. Zalewski, Z. Phys. C{\bf 23}  (1984) 233.

\bibitem{VC90} G. Mennessier and B. Causse, Z. Phys. C{\bf 47}  (1990) 611; S. Narison, Phys. Lett. B{\bf 387} (1996) 162; F. J. Yndurain, Phys. Rept. {\bf 320} (1999) 287.

\bibitem{VC20} B. L. Ioffe and K. N. Zyablyuk, Eur. Phys. J. C{\bf 27} (2003) 229.

\bibitem{REW} P. Colangelo and A. Khodjamirian, hep-ph: 0010175.

\bibitem{VC2010} S. Narison, arXiv:1105.2922 [hep-ph]; arXiv: 1010.1959[hep-ph].

\bibitem{VC2011} S. Narison, arXiv:1105.5070 [hep-ph].

\bibitem{as} S. Bethke, Eur. Phys. J. C{\bf 64} (2009) 689.

\bibitem{rge90} K. C. Yang and W-Y. P. Hwang, Phys. Rev. D{\bf 47} (1993) 3001; W-Y. P. Hwang and K. C. Yang, Phys. Rev. D{\bf 49} (1994) 460.

\bibitem{rge20} C.D. Lu, Y.M. Wang, and H. Zou, Phys. Rev. D{\bf 75} (2007) 056001.

\bibitem{nlomu} L.J. Reinders, H.R. Rubinstein, and S. Yazaki, Phys.Rept.{\bf 127} (1985) 1.

\bibitem{vac_con2} B. L. Ioffe and A. V. Smilga, Nucl. Phys. B{\bf 216} (1983) 373.

\bibitem{vac_con_LJW} D.S. Du, J.W. Li, and M.Z. Yang, Eur. Phys. J. C{\bf 37} (2004) 173.

\end{thebibliography}
\end{document}